\documentclass[10pt,twocolumn]{elsart5p}
\usepackage{graphicx}
\usepackage{dcolumn}
\usepackage{amsmath}
\usepackage{amssymb}
\usepackage{bm}
\usepackage{natbib}

\begin{document}

\begin{frontmatter}

\title{Monitoring spike train synchrony}

\author[ISC]{Thomas Kreuz},
\ead{thomas.kreuz@cnr.it}
\author[ITT]{Daniel Chicharro},
\author[UBE]{Conor Houghton},
\author[UPF]{Ralph G. Andrzejak},
\author[UBG]{Florian Mormann}

\address[ISC]{Institute for complex systems, CNR, Sesto Fiorentino, Italy}
\address[ITT]{Center for Neuroscience and Cognitive Systems@UniTn, Italian Institute of Technology, Rovereto (TN), Italy}
\address[UBE]{Department of Computer Science, University of Bristol, Bristol, England}
\address[UPF]{Department of Information and Communication Technologies, Universitat Pompeu Fabra, Barcelona, Spain}
\address[UBG]{Department of Epileptology, University of Bonn, Bonn, Germany}

\date{\today}

\begin{abstract}

Recently, the SPIKE-distance has been proposed as a parameter-free and time-scale independent measure of spike train synchrony. This measure is time-resolved since it relies on instantaneous estimates of spike train dissimilarity. However, its original definition led to spuriously high instantaneous values for event-like firing patterns. Here we present a substantial improvement of this measure which eliminates this shortcoming. The reliability gained allows us to track changes in instantaneous clustering, i.e., time-localized patterns of (dis)similarity among multiple spike trains. Additional new features include selective and triggered temporal averaging as well as the instantaneous comparison of spike train groups. In a second step, a causal SPIKE-distance is defined such that the instantaneous values of dissimilarity rely on past information only so that time-resolved spike train synchrony can be estimated in real-time. We demonstrate that these methods are capable of extracting valuable information from field data by monitoring the synchrony between neuronal spike trains during an epileptic seizure. Finally, the applicability of both the regular and the real-time SPIKE-distance to continuous data is illustrated on model electroencephalographic (EEG) recordings.

\end{abstract}


\end{frontmatter}

\newcommand{\abb}{\small\sf}
\maketitle

%
%
\section{\label{s:Intro} Introduction}

Measures that estimate the degree of synchrony between two or more simultaneously recorded spike trains are important tools in many different areas of neuroscientific research \citep{Kreuz11b}. Among many potential applications they can be used to evaluate the role of synchronous neuronal firing in signal propagation \citep{Reyes03}, to estimate the pairwise correlation within a neuronal population \citep{Pillow08}, or to quantify the role of spike synchronization in feature binding \citep{Singer09}.

Some of the most widely used measures of spike train dissimilarity depend on a parameter that determines the temporal scale in the spike trains to which the measure is sensitive. Examples include the Victor-Purpura spike train distance \citep{Victor96} and the van Rossum distance \citep{VanRossum01}. Recently, the ISI-distance \citep{Kreuz07c, Kreuz09} and the SPIKE-distance \citep{Kreuz11} have been proposed as parameter-free and time-scale adaptive alternatives. These two measures are complementary to each other: While the ISI-distance quantifies local dissimilarities based on the neurons' firing rate profiles, the SPIKE-distance tracks dyssynchrony mediated by differences in spike timing. The latter kind of sensitivity can be very relevant since coincident spiking is found in many different neuronal circuits, e.g., in the visual cortex \citep{Usrey99, Priebe08} and in the retina \citep{Meister99, Shlens08}.

In some situations it is sufficient to evaluate spike train synchrony at a rather low temporal resolution, e.g., by means of a moving-window analysis where the level of synchronization within a certain interval is compressed into a single value and then compared for successive intervals. On the other hand, many applications require a high temporal resolution, e.g., in order to detect replay of precisely timed sequential patterns of neural activity \citep{Ji07}, to track spike train response variability within a neuronal population \citep{Mitchell07, Kreuz09}, or to understand the role of synchronous firing in the neuronal coding of time-dependent stimuli \citep{Miller08}. In epilepsy research, a high temporal resolution could help to gain a deeper understanding of the neuronal spiking patterns involved in the different phases of seizure generation, propagation, and termination \citep{Truccolo11, Bower12}.

Previously, the Victor-Purpura and the van Rossum distance (as well as some other spike train distances) were compared against the ISI- \citep{Kreuz07c, Kreuz09} and the SPIKE-distance \citep{Kreuz11} regarding their capability to evaluate the overall similarity of two or more spike trains. However, the maximum possible temporal resolution is achieved when one value of dissimilarity is calculated for each time instant and a continuous time profile is obtained. In contrast to the Victor-Purpura and the van Rossum distance, the ISI- and the SPIKE-distance can be calculated and visualized in such a time-resolved manner. Yet, as already noted in \citet{Kreuz11}, while the original definition of the SPIKE-distance yields the expected results after temporal averaging and also correctly reflects long-term trends by means of a moving average, slightly unreliable spiking events lead to spurious high instantaneous values. In the first part of this study, we remedy this problem by considering spike time differences between nearest spikes instead of separating differences between preceding spikes and differences between following spikes.

After improving the SPIKE-distance, we extend its applicability to situations where the degree of synchrony between two or more simultaneously recorded spike trains is monitored in real-time. In the field of brain-machine interfaces this may be a promising approach to the rapid online decoding of neural signals needed to control prosthetics \citep{Hochberg06, Sanchez08}. In epileptic patients who are refractory to medical treatment, the method could be applied to large ensembles of single neurons close to or within the epileptic focus and then be integrated into a prospective algorithm aimed, e.g., at an early seizure detection \citep{Jouny11}.

Most spike train distances calculate one value of overall spike train synchrony for a time interval once this time interval has passed. There is a lack of measures that can estimate synchrony in a time-resolved and, even more striking, in a causal way. The SPIKE-distance, similar to the ISI-distance, is calculated from instantaneous values of spike train dissimilarity for which at each time moment not only preceding spikes but also following spikes are taken into account. This non-causal dependence on 'future' spiking does not allow for a real-time calculation. In the second part of this study we modify the SPIKE-distance such that the instantaneous value of dissimilarity for two or more spike trains relies on past information only and can be calculated in real-time and in a causal manner.

We illustrate the new methods on three different applications. First we validate the improved definition of the SPIKE-distance and its real-time variant on artificially generated spike trains. We show that these measures are not only able to track the overall synchronization within a group of two or more spike trains, but also, due to the reliability added by the revised definition of the distance, to track and visualize changes in instantaneous clustering, that is, to follow the evolution of (dis)similarity patterns within multiple spike trains. Furthermore, we demonstrate additional features such as selective and (internally and externally) triggered temporal averaging as well as the comparison of spike train groups. Subsequently, we present an application to real spike train data in which we analyze single- and multi-unit activity from an epilepsy patient recorded in a time interval which includes the occurrence of an epileptic seizure. We can show that the new spike train distances are suitable measures to characterize the neuronal firing patterns involved in seizure generation, propagation, and termination.

Finally, as for spike trains, there is a lack of measures that are capable to monitor time-resolved synchrony in continuous data. This issue is addressed in the third and last application in which both variants of the SPIKE-distance are used to measure the time-resolved dissimilarity in continuous data which, in a preprocessing step, are first transformed into discrete data. We illustrate this approach on electroencephalogram (EEG) time series whose level of synchronization was modified post hoc in a controlled manner by means of linear mixing.
%
%
\section{\label{s:Methods} Methods}

The different variants of the SPIKE-distance, and the ISI-distance (see Appendix \ref{App-s:Bivariate-ISI-Distance}), are defined in terms of a function of time, a time profile for each pair of spike trains which gives an instantaneous measure of the (dis)similarity between the two spike trains. The distances are then defined as the temporal average of these dissimilarity profiles, e.g., for the bivariate SPIKE-distance,
\begin{equation} \label{eq:Temporal-Average}
    D_S = \frac{1}{T} \int_{t=0}^T S (t) dt
\end{equation}
where $T$ denotes the overall length of the spike trains, e.g., the duration of the recording in an experiment. In the following this equation is always omitted, and we restrict ourselves to showing how to derive the respective dissimilarity profiles, e.g., $S (t)$.

Furthermore, for all distances there exists a straightforward extension to the case of more than two spike trains (number of spike trains $N > 2$), the averaged bivariate distance. This average over all pairs of neurons commutes with the average over time, so it is possible to achieve the same kind of time-resolved visualization as in the bivariate case by first calculating the instantaneous average, e.g., $S^{\mathrm {a}} (t)$ over all pairwise instantaneous values $S^{mn} (t)$,
\begin{equation} \label{eq:Bivariate-Average}
    S^{\mathrm {a}} (t) = \frac{1}{N(N-1)/2}\sum_{n=1}^{N-1} \sum_{m=n+1}^N S^{mn} (t)
\end{equation}
and only then averaging the resulting dissimilarity profile using Eq. \ref{eq:Temporal-Average}. All bivariate and averaged bivariate dissimilarity profiles and thus all distances are bounded in the interval $[0, 1]$. The value zero is only obtained for identical spike trains.

\subsection{\label{ss:Methods-11-Old-Spike-Distance} Original definition of the SPIKE-distance}

We first review the original definition of the time-resolved profile for the bivariate SPIKE-distance \citep{Kreuz11}. This profile is denoted as $S_o(t)$ with the subscript \textquoteleft$o$' standing for original. We then compare it against the improved profile, denoted as $S (t)$, by which it will henceforth be replaced. The derivation consists of three steps: calculating the instantaneous time differences between spikes, taking the locally weighted average and normalizing the result.

\begin{figure}
    \includegraphics[width=85mm]{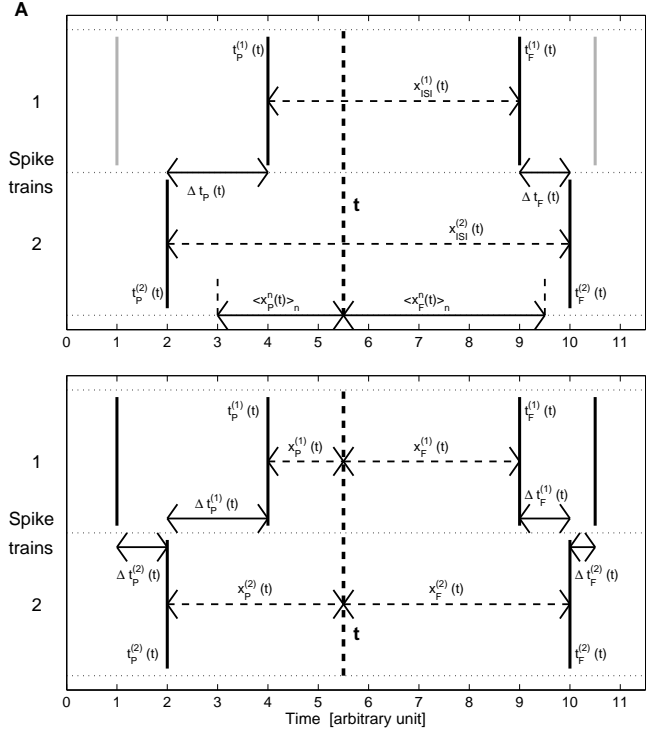}
    \caption{\abb\label{fig:Fig1-Illustration} Illustration of the bivariate SPIKE-distance.   A. Local quantities in relation to time instant $t$ needed to define the original dissimilarity profile (Eq. \ref{eq:Bi-Spike-Diss-Original}) of the SPIKE-distance.   B. Additional definitions used for the improved dissimilarity profile (Eqs. \ref{eq:Bi-Spike-Diss-Improved-First} and \ref{eq:Bi-Spike-Diss-Improved}) and its real-time variant (Eq. \ref{eq:Bi-Spike-Diss-RT}).}
\end{figure}

After labeling the times of the spikes in the spike trains $n = 1, 2$ by $t_i^{(n)}, i=1,...,M_n$ (with $M_n$ denoting the number of spikes of the $n$-th spike train), we assign to each time instant between $0$ and $T$ (see Fig. \ref{fig:Fig1-Illustration}A) the time of the preceding spikes
\begin{equation} \label{eq:Prev-Spike}
    t_{\mathrm {P}}^{(n)} (t) = \max_i (t_i^{(n)} | t_i^{(n)} \leq t),
\end{equation}
the time of the following spikes
\begin{equation} \label{eq:Foll-Spike}
    t_{\mathrm {F}}^{(n)} (t) = \min_i (t_i^{(n)} | t_i^{(n)} > t),
\end{equation}
as well as the instantaneous interspike interval
\begin{equation} \label{eq:ISI}
    x_{\mathrm {ISI}}^{(n)} (t) = t_{\mathrm {F}}^{(n)} (t) - t_{\mathrm {P}}^{(n)} (t).
\end{equation}

The ambiguity regarding the definition of the very first and the very last interspike interval as well as the initial distance to the preceding spike and the final distance to the following spike is resolved by adding auxiliary leading spikes at time $t = 0$ and auxiliary trailing spikes at time $t = T$ to each spike train.

We then denote the instantaneous absolute differences of preceding and following spike times as
\begin{equation} \label{eq:Prev-Diff}
    \Delta t_{\mathrm {P}} (t) = | t_{\mathrm {P}}^{(1)} (t) - t_{\mathrm {P}}^{(2)} (t) |
\end{equation}
and
\begin{equation} \label{eq:Foll-Diff}
    \Delta t_{\mathrm {F}} (t) = | t_{\mathrm {F}}^{(1)} (t) - t_{\mathrm {F}}^{(2)} (t) |,
\end{equation}
respectively. An instantaneous spike time based measure of spike train distance is given by the average of these two absolute differences of preceding and following spike times. Dividing this average spike time difference by the average of the two instantaneous interspike intervals achieves proper normalization as well as time-scale invariance (i.e., stretching or compressing spike trains does not change the result):
\begin{equation} \label{eq:Bi-Spike-Diss-Prelim}
    S' (t) = \frac{\Delta t_{\mathrm {P}} (t) + \Delta t_{\mathrm {F}} (t)}{2 \langle x_{\mathrm {ISI}}^{(n)} (t) \rangle_n}.
\end{equation}

In order to be more local in time, a weighted average of the two differences $\Delta t_{\mathrm {j}} (t)$ with $\mathrm {j} = \mathrm {P},\mathrm {F}$ is employed such that the difference of the spikes that are closer dominate. To this aim we denote with
\begin{equation} \label{eq:Prev-Spike-Dist}
     x_{\mathrm {P}}^{(n)} (t) = t - t_{\mathrm {P}}^{(n)} (t)
\end{equation}
and
\begin{equation} \label{eq:Foll-Spike-Dist}
     x_{\mathrm {F}}^{(n)} (t) = t_{\mathrm {F}}^{(n)} (t) - t
\end{equation}
the intervals to the previous and the following spikes for each neuron $n = 1, 2$.

Inserting the inverse of the mean intervals as weights in the locally weighted average
\begin{equation} \label{eq:Weighted-Mean}
    \langle \Delta t_{\mathrm {j}} (t) \rangle_{\mathrm {j}=\mathrm {P},\mathrm {F}} = \frac{\sum_{\mathrm {j}=\mathrm {P},\mathrm {F}} \Delta t_{\mathrm {j}} (t) \frac{1}{\langle x_{\mathrm {j}}^{(n)} (t) \rangle_n}}{\sum_{\mathrm {j}=\mathrm {P},\mathrm {F}} \frac{1}{\langle x_{\mathrm {j}}^{(n)} (t) \rangle_n}}
\end{equation}
and making use of
\begin{equation} \label{eq:Multi-Mean-Foll}
     x_{\mathrm {P}}^{(n)} (t) + x_{\mathrm {F}}^{(n)} (t) = x_{\mathrm {ISI}}^{(n)} (t)
\end{equation}
yields the dissimilarity profile
\begin{equation} \label{eq:Bi-Spike-Diss-Original}
     S_o (t) = \frac{\Delta t_{\mathrm {P}} (t) \langle x_{\mathrm {F}}^{(n)} (t) \rangle_n + \Delta t_{\mathrm {F}} (t) \langle x_{\mathrm {P}}^{(n)} (t) \rangle_n}
        {\langle x_{\mathrm {ISI}}^{(n)} (t) \rangle_n^2}.
\end{equation}
For multiple spike trains the averaged bivariate variant is defined via Eq. \ref{eq:Bivariate-Average}.

\subsection{\label{ss:Methods-13-New-Spike-Distance} Improved definition of the SPIKE-distance}

In its original definition in \citet{Kreuz11}, the SPIKE-distance proved to be a reliable indicator of the overall level of multi-neuron synchrony for both simulated and real data. Appropriate moving averaging also allowed to correctly reflect long-term changes in spike train synchrony. However, as already noted in \citet{Kreuz11}, each individual instantaneous value in itself is less reliable since spuriously high values are obtained during slightly unreliable spiking events.

\begin{figure}
    \includegraphics[width=85mm]{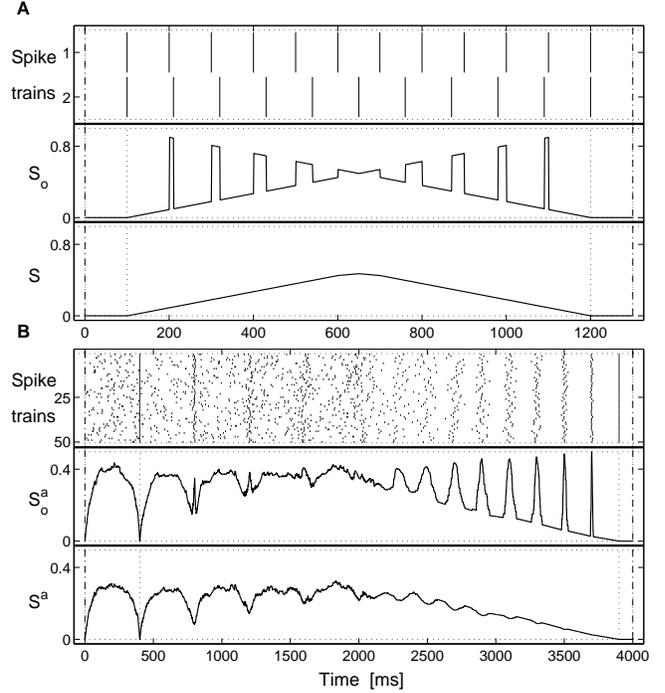}
    \caption{\abb\label{fig:Fig2-Bi-Multi-Comparison} Comparison of the original and the improved definition of the dissimilarity profile of the SPIKE-distance on constructed spike trains. While $S_o (t)$ shows spurious high values for event-like firing patterns, $S (t)$ reflects the level of spike train synchrony faithfully.   A. Bivariate example: Varying spike matches. B. Multivariate example with $50$ spike trains: In the first half within the noisy background there are four regularly spaced spiking events with increasing jitter. The second half consists of ten spiking events with decreasing jitter but now without any noisy background. Note that both dissimilarity profiles start at zero due to the auxiliary spikes at time $t=0$.}
\end{figure}

This can be seen in the bivariate example of Fig. \ref{fig:Fig2-Bi-Multi-Comparison}A where we use a frequency mismatch to construct two spike trains with gradually varying spike matches. In the first half the spikes in the second spike train exhibit increasing distances to the preceding spikes of the first spike train, while in the second half they move closer and closer to its following spikes. Thus we expect a monotonic increase of the instantaneous values followed by a monotonic decrease. These two trends can indeed be recognized. However they are interrupted by short intervals of spurious high values.

The same kind of spurious high values can be seen in Fig. \ref{fig:Fig2-Bi-Multi-Comparison}B for the averaged bivariate variant $S_o^a (t)$ in a multivariate example. Here in the first half we generated four spiking events with increasing jitter within a noisy background whereas the second half consists of evenly spaced firing events with increasing precision, this time without any background noise. In both cases the spurious high values appear in small intervals during which some of the spikes are still following spikes while others are already preceding spikes. In these intervals the small differences between the spikes within the event are not taken into account. Instead, due to the separation of preceding and following spikes the 'wrong' spikes are compared to each other. Some of the differences among the preceding spikes as well as among the following spikes are very large and this, according to Eq. \ref{eq:Bi-Spike-Diss-Original}, leads to the high instantaneous values.

A straightforward remedy for this mismatch is to compare the correct spikes to each other, in this case, to compare each spike to the nearest spike in the other spike train. To do so we first rewrite $S_o (t)$. For the sake of brevity we now omit time dependencies. Using $2\langle x_{\mathrm {j}}^{(n)} \rangle_n= x_{\mathrm {j}}^{(1)} + x_{\mathrm {j}}^{(2)}$ with  $\mathrm{j} = \mathrm{P},\mathrm {F}$ gives
\begin{eqnarray} \label{eq:Bi-Spike-Diss-Original-Rewritten1}
     S_o (t) & = & \frac{\Delta t_{\mathrm {P}} \langle x_{\mathrm {F}}^{(n)} \rangle_n + \Delta t_{\mathrm {F}} \langle x_{\mathrm {P}}^{(n)} \rangle_n}{\langle x_{\mathrm {ISI}}^{(n)} \rangle_n^2} \\ \nonumber
        & = & \frac{ \Delta t_{\mathrm {P}} x_{\mathrm {F}}^{(1)} + \Delta t_{\mathrm {P}} x_{\mathrm {F}}^{(2)} +  \Delta t_{\mathrm {F}} x_{\mathrm {P}}^{(1)} + \Delta t_{\mathrm {F}} x_{\mathrm {P}}^{(2)}}{2 \langle x_{\mathrm {ISI}}^{(n)} \rangle_n^2}
\end{eqnarray}
and expanding the definition of $\Delta t_{\mathrm{P}}$ and $\Delta t_{\mathrm{F}}$ (Eqs. \ref{eq:Prev-Diff} and \ref{eq:Foll-Diff}) yields
\begin{eqnarray} \label{eq:Bi-Spike-Diss-Original-Rewritten}
     S_o(t)   & = & \frac{ | t_{\mathrm {P}}^{(1)} - t_{\mathrm {P}}^{(2)} | x_{\mathrm {F}}^{(1)} + | t_{\mathrm {P}}^{(2)} - t_{\mathrm {P}}^{(1)} | x_{\mathrm {F}}^{(2)}}{2 \langle x_{\mathrm {ISI}}^{(n)} \rangle_n^2} +  \\
        & & \frac{| t_{\mathrm {F}}^{(1)} - t_{\mathrm {F}}^{(2)} | x_{\mathrm {P}}^{(1)} + | t_{\mathrm {F}}^{(2)} - t_{\mathrm {F}}^{(1)} | x_{\mathrm {P}}^{(2)}}{2 \langle x_{\mathrm {ISI}}^{(n)} \rangle_n^2}. \nonumber
\end{eqnarray}
In this formulation $S_o$ can be interpreted as the normalized sum of four weighted differences, one each for the preceding spike from the first spike train $t_{\mathrm {P}}^{(1)}$, the following spike from the first spike train $t_{\mathrm {F}}^{(1)}$, the preceding spike from the second spike train $t_{\mathrm {P}}^{(2)}$, and, finally, the following spike from the second spike train $t_{\mathrm {F}}^{(2)}$. However, as we have established above, in some instances these four \emph{corner spikes} are compared against the 'wrong' spikes. This happens because we always restrict the spikes of comparison to the respective other preceding and to the respective other following spike.

A way to resolve this restriction is to allow more flexibility and compare each of these four corner spikes to its most appropriate spike, i.e., the closest counterpart in the other spike train. To this aim we define
\begin{equation} \label{eq:Delta-Corner-Spike}
     \Delta t_{\mathrm {P}}^{(1)} = \min_i (| t_{\mathrm {P}}^{(1)} - t_i^{(2)} |)
\end{equation}
and analogously for $t_{\mathrm {F}}^{(1)}$, $t_{\mathrm {P}}^{(2)}$, and $t_{\mathrm {F}}^{(2)}$ (see Fig. \ref{fig:Fig1-Illustration}B). In the improved dissimilarity profile these four terms replace the twofold contributions of $|\Delta t_{\mathrm {P}}|$ and $|\Delta t_{\mathrm {F}}|$. Furthermore, instead of one local weighted average of the two differences between previous and following spikes with the mean intervals to the previous and the following spikes as weights (Eq. \ref{eq:Weighted-Mean}), the weighting is now carried out for each spike train separately. The local weighting for the spike time differences of the first spike train reads
\begin{eqnarray} \label{eq:Bi-Spike-Diss-Improved-First}
     S_1 (t) & = & \frac{\frac{\Delta t_{\mathrm {P}}^{(1)}}{x_{\mathrm {P}}^{(1)}} + \frac{\Delta t_{\mathrm {F}}^{(1)}}{x_{\mathrm {F}}^{(1)}}}{\frac{1}{x_{\mathrm {P}}^{(1)}} + \frac{1}{x_{\mathrm {F}}^{(1)}}} \\
     & = & \frac{\Delta t_{\mathrm {P}}^{(1)} x_{\mathrm {F}}^{(1)} + \Delta t_{\mathrm {F}}^{(1)} x_{\mathrm {P}}^{(1)}}{x_{\mathrm {F}}^{(1)} + x_{\mathrm {P}}^{(1)}} \nonumber \\
     & = & \frac{\Delta t_{\mathrm {P}}^{(1)} x_{\mathrm {F}}^{(1)} + \Delta t_{\mathrm {F}}^{(1)} x_{\mathrm {P}}^{(1)}}{x_{\mathrm {ISI}}^{(1)}} \nonumber
\end{eqnarray}
and analogously $S_2 (t)$ is obtained for the second spike train. Averaging over the two spike train contributions and normalizing by the mean interspike interval yields
\begin{equation} \label{eq:Bi-Spike-Diss-Improved-Intermediate}
     S'' (t) = \frac{S_1 (t) + S_2 (t)}{2 \langle x_{\mathrm {ISI}}^{(n)} \rangle_n}.
\end{equation}

This quantity weights the spike time differences for each spike train according to the relative distance of the corner spike from the time instant under investigation. This way relative distances within each spike train are taken care of, while relative distances between spike trains are not. In order to get these ratios straight and to account for differences in firing rate, in a last step the two contributions from the two spike trains are locally weighted by their instantaneous interspike intervals. This leads to the improved definition of the dissimilarity profile
\begin{equation} \label{eq:Bi-Spike-Diss-Improved}
     S (t) = \frac{S_1 (t) x_{\mathrm {ISI}}^{(2)} + S_2 (t) x_{\mathrm {ISI}}^{(1)}}{2 \langle x_{\mathrm {ISI}}^{(n)} \rangle_n^2}.
\end{equation}

For many time instants the closest spike to the preceding spike of one spike train is the preceding spike in the other spike train and the same holds for the following spikes, so often the same spikes are still compared to each other. However, for the time instants that led to the spurious high values for $S_o (t)$ now the two preceding and the two following spikes are no longer compared to each other. Instead if a following spike in one spike train is the closest spike to a preceding spike of the other spike train (or vice versa) these will now be compared to each other, which leads to lower instantaneous values in the dissimilarity profile.

This modification resolves the problem of spurious high values and, as can be seen in Fig. \ref{fig:Fig2-Bi-Multi-Comparison}, the desired dissimilarity profiles are obtained. In Fig. \ref{fig:Fig2-Bi-Multi-Comparison}A the observed monotonic increase of the instantaneous values followed by a monotonic decrease mirrors exactly the actual change in the match between the spike times of the two spike trains. For the multivariate example of Fig. \ref{fig:Fig2-Bi-Multi-Comparison}B and the averaged bivariate variant (calculated according to Eq. \ref{eq:Bivariate-Average}) we find in the first half high instantaneous values which reflect the background noise. The presence of four spiking events with increasing jitter within this noisy background is indicated by less and less pronounced drops in the dissimilarity profile. In the second half where there is no background noise, the evenly spaced firing events with increasing precision are correctly reflected by a rather monotonic decrease of the instantaneous synchrony level which reaches zero at the perfectly synchronous event at $3900$ ms. Such perfect events for which the value zero is obtained are marked by vertical dashed lines in the dissimilarity profile.

The firing rate correction introduced between Eqs. \ref{eq:Bi-Spike-Diss-Improved-Intermediate} and \ref{eq:Bi-Spike-Diss-Improved} has the important and desirable consequence that the SPIKE-distance between Poisson spike trains increases with the difference in firing rate (results not shown).
%
%
\subsection{\label{ss:Methods-14-Real-time-Spike-Distance} Real-time SPIKE-distance}

Here we introduce the real-time SPIKE-distance $D_{S_r}$. This is a modification of the SPIKE-distance with the key difference that the corresponding time profile $S_r(t)$ can be calculated online because it relies on past information only. From the perspective of an online measure, the information provided by the following spikes, both their position and the length of the interspike interval, is not yet available. Like the regular (improved) SPIKE-distance $D_S$, this causal variant is also based on local spike time differences but now only two corner spikes are available, and the spikes of comparison are restricted to past spikes, e.g., for the preceding spike of the first spike train
\begin{equation} \label{eq:Delta-Corner-Spike-Real-time}
     \Delta t_{\mathrm {P}}^{(1)} = \min_i (| t_{\mathrm {P}}^{(1)} - t_i^{(2)} |), t_i < t.
\end{equation}
Since there are no following spikes available, there is no local weighting, and since there is no interspike interval, the normalization is achieved by dividing the average corner spike difference by twice the average time interval to the preceding spikes (Eq. \ref{eq:Prev-Spike-Dist}, see also Fig. \ref{fig:Fig1-Illustration}B). This yields a causal indicator of local spike train dissimilarity:
\begin{equation} \label{eq:Bi-Spike-Diss-RT}
    S_r (t) = \frac{ \Delta t_{\mathrm {P}}^{(1)} + \Delta t_{\mathrm {P}}^{(2)}} {4 \langle x_{\mathrm {P}}^{(n)} \rangle_n}.
\end{equation}

\begin{figure}
    \includegraphics[width=85mm]{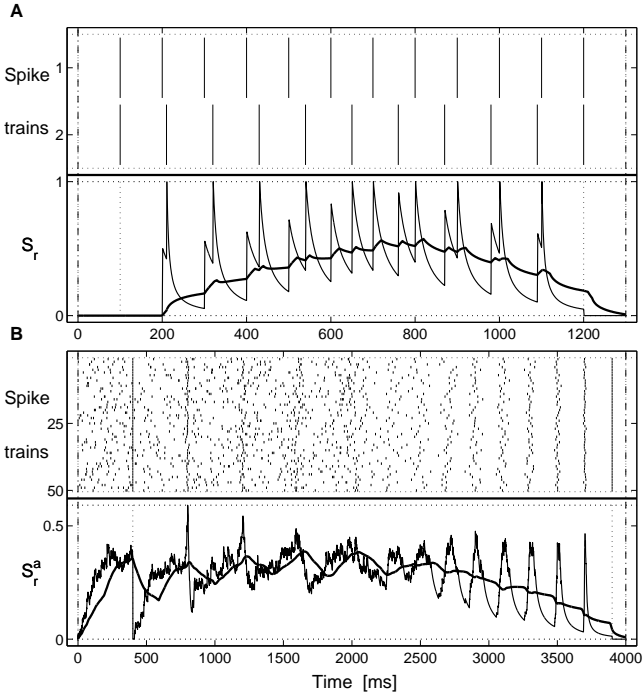}
    \caption{\abb\label{fig:Fig3-Bi-Multi-Comparison-Realtime-Comp} Time profile of the real-time SPIKE-distance for the two examples used in Fig. \ref{fig:Fig2-Bi-Multi-Comparison}. In both subplots regular traces are depicted in black while the causal moving averages are shown in green.    A. Bivariate example. While the regular dissimilarity profile $S_r (t)$ attains a local maximum for each spike, its subsequent decay still captures the relative spiking behavior. This can be seen best with the moving average which reflects the correct long-term trends.    B. Multivariate example. The averaged bivariate dissimilarity profile $S_r^a (t)$ exhibits periods of high values in certain intervals, but in contrast to the original non-causal case $S_o^a (t)$, these are not spurious (see Fig. \ref{fig:Fig4-Not-Spurious}). }
\end{figure}

In Fig. \ref{fig:Fig3-Bi-Multi-Comparison-Realtime-Comp} we show the results of the real-time SPIKE-distance for the two cases already used in Fig. \ref{fig:Fig2-Bi-Multi-Comparison}. As can be seen for the example of two spike trains with gradually varying spike mismatches (Fig. \ref{fig:Fig3-Bi-Multi-Comparison-Realtime-Comp}A), any spike time difference is considered most relevant right at the later of two spikes when $S_r (t)$ goes back to a local maximum value. In the case where the two preceding spikes are closest to each other, it goes back to its maximum value of one. At these points the mean time interval to the two preceding spikes is exactly half their difference. Any successive period of common non-spiking leads to a decrease of the instantaneous distance values. This is a desired property since common non-spiking is as much a sign of synchrony as common spiking. The decrease is hyperbolic and its slope depends on the preceding spike time difference.

In addition to the regular trace we here also show a moving average which, in line with the real-time calculation, is causal. While the regular dissimilarity profile exhibits certain fluctuations, this moving average shows that the real-time SPIKE-distance in fact reflects the increasing spike shifts in the center of the interval and the better match of the spike times at the edges.

During irregular spiking in the multivariate example of Fig.\ref{fig:Fig3-Bi-Multi-Comparison-Realtime-Comp}B, the time intervals to the preceding spikes in different spike trains are very variable, and this leads to large fluctuating values of $S_r^{\mathrm {a}} (t)$. Within this irregularity, the perfect spiking event at $400$ ms results in an abrupt drop to zero, which reflects the delta distribution of the time intervals to the preceding spikes. The successively more jittered events which follow are indicated by a very pronounced short-term increase, followed by a decrease. Here the distribution of time intervals to preceding spikes starts to develop a peak at very small intervals and thus becomes bimodal, which causes the increase. Then, after spikes have appeared in quick succession in all spike trains, the distribution becomes very narrow, which is reflected by the decrease. In the second half, when there is no background noise and the spiking events become less and less noisy, the succession of peaks denoting the events is becoming more and more prominent with increasing amplitude range but narrower base. Just as in the bivariate case, the short time-scale fluctuations in this multivariate example can be eliminated by an appropriate (causal) moving average. Here this moving average exhibits a gradual decrease, which reflects the consistent increase of the spike event reliability.
\begin{figure}
    \includegraphics[width=85mm]{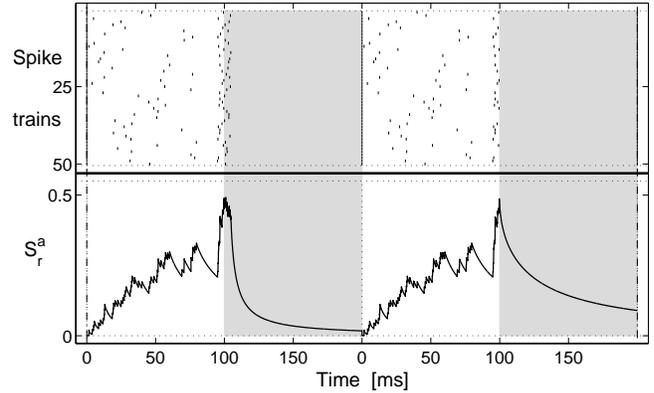}
    \caption{\abb\label{fig:Fig4-Not-Spurious} Real-time SPIKE-distance: Peaks during reliable spiking events are not spurious. Multivariate example with two spiking events which have identical first halves but different continuations. After $100$ ms some of the neurons have just spiked while others have been silent for some time, and for both events this large variability is correctly reflected by high values of the averaged bivariate dissimilarity profile $S_r^a (t)$. At these points begin the shaded areas which mark the spike information that is not yet available. Here for the first event the initial spiking in some of the spike trains turns into a reliable spiking event (one spike for each spike train) which is reflected by a rapid decrease to very low values. For the second event this does not happen and accordingly the decrease is less pronounced.}
\end{figure}

It is important to note that, in contrast to the non-causal SPIKE-distance (see Fig. \ref{fig:Fig2-Bi-Multi-Comparison}), the observed peaks are not spurious. They occur at time instants when it is not yet known whether there will in fact be a reliable spiking event or not. This is illustrated in Fig. \ref{fig:Fig4-Not-Spurious} with two spiking events which are identical (one spike per spike train) except for the omitted second half of the second event. While for both events the instantaneous values in the first part necessarily have to be identical (and very high since only some of the neurons have recently spiked), the differences in the second part become evident as more and more information becomes available. For the first event, once all neurons have fired, a rapid decrease of $S_r^a (t)$ can be observed, while the lower firing reliability in the second event leads to a slow decrease.

\subsection{\label{ss:Practical-considerations} Practical considerations}

The improved dissimilarity profile $S (t)$ of the SPIKE-distance is piecewise linear (with each linear interval running from one spike of the pooled spike train to the next) rather than piecewise constant as is the case for the ISI-distance. Therefore, when the localized visualization is desired, a new value has to be calculated for each sampling point and not just once per each interval in the pooled spike train. In cases where the distance value itself is sufficient, the short computation time can be even further decreased by representing each interval by the value of its center and weighting it by its length. This is not only faster, but it actually gives the exact result, whereas the time-resolved calculation is a very good approximation only for sufficiently small sampling intervals $dt$ (imagine the example of a rectangular function, at some point any sampled representation has to cut the right angle). The dissimilarity profile $S_r (t)$ of the real-time SPIKE-distance is hyperbolic and not linear but here also the exact result can be obtained by piecewise integration over all intervals of the pooled spike train.

The calculation of the SPIKE-distance consists of three steps: In a precalculation step, for each spike the distance to the nearest spike in all the other spike trains is calculated. Successively, for each time instant and each pair of spike trains, the distances of the four corner spikes are first locally weighted and then normalized. These latter steps involve matrices of the order 'number of time instants' $\times$ 'number of spike train pairs', which for very long datasets with many spike trains can lead to memory problems. The solution to these problems is to make the calculation sequential, i.e., to cut the recording interval into smaller segments, and to perform the averaging over all pairs of spike trains for each segment separately (no additional auxiliary spikes are needed except for very huge datasets for which even the calculation of the first matrix is too memory-demanding). In the end the dissimilarity profiles for the different segments (already averaged over pairs of spike trains) are concatenated, and its temporal average yields the distance value for the whole recording interval.

The computational load scales with the number of spike trains $N$ as $N^2$. Using Matlab on a notebook with a 2.53GHz Intel Core 2 Duo processor the calculation of most of the examples in this article took just a few seconds, while the slowest one (Fig. 9) took less than 5 minutes.

More information on the implementation as well as the Matlab source code for both the ISI- and the SPIKE-distance (including its real-time variant) can be found at www.fi.isc.cnr.it/users/thomas.kreuz/sourcecode.html.

%
\section{\label{s:Results} Results}

\subsection{\label{ss:Application-Clustering} Application to artificially generated data: Instantaneous clustering}

By eliminating the spurious high values in $S_o^a (t)$, we have gained reliability of $S^a (t)$ for each time instant. This allows us to use the instantaneous matrices of pairwise spike train dissimilarities to divide the spike trains into clusters, i.e., groups of spike trains with low intra-group and high inter-group dissimilarity. There are no limits to the temporal resolution; such a clustering can, in principle, be obtained for each time instant.
\begin{figure}
    \includegraphics[width=85mm]{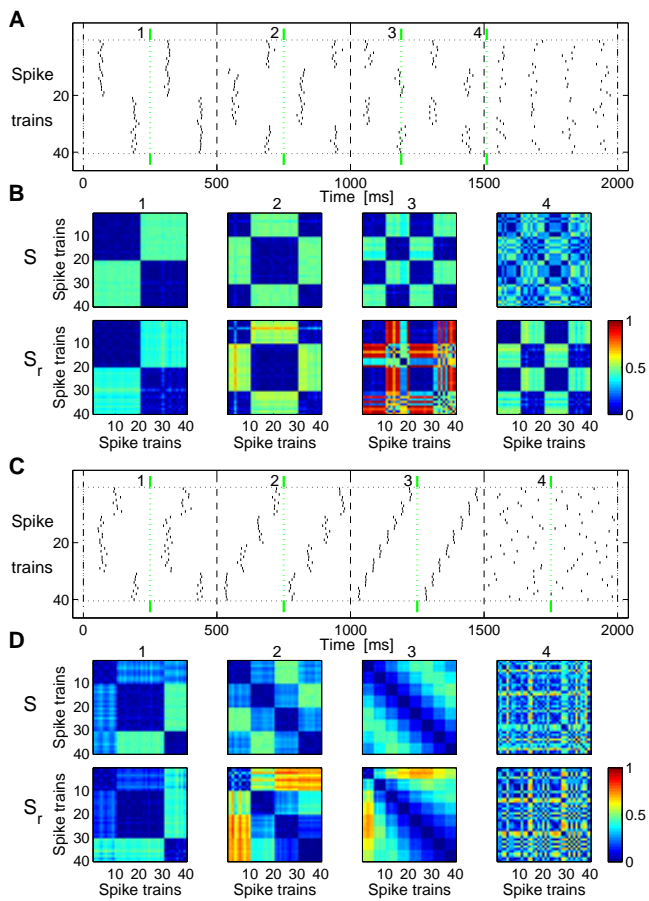}
    \caption{\abb\label{fig:Fig5-Instantaneous-Clustering} Instantaneous clustering for artificially generated spike trains.   A. The clustering of the spike trains changes every $500$ ms (dashed black lines) although there are always exactly two different clusters.   B. Matrices of pairwise instantaneous values for the four time instants marked by the green lines. Both the regular (top) and the real-time SPIKE-distance (bottom) reflect the changing cluster association in each $500$ ms interval. Arrows indicate features described in the text.   C. Here the three spike train clusters in the first $500$ ms are followed by $500$-ms-intervals of four and then eight clusters while the last $500$ ms contain random spiking.   D. Again, both the regular (top) and the real-time SPIKE-distance (bottom) reflect changes in instantaneous clustering.}
\end{figure}

In Fig. \ref{fig:Fig5-Instantaneous-Clustering} we show examples of instantaneous clustering for both the regular and the real-time SPIKE-distance. Both figures depict artificially generated spike trains which fall into different clusters. However, in contrast to the overall level of spike train synchrony which remains rather constant (results not shown), the cluster affiliation of the different spike trains changes every $500$ ms. The spike trains in Fig. \ref{fig:Fig5-Instantaneous-Clustering}A contain four different compositions of two clusters, whereas in Fig. \ref{fig:Fig5-Instantaneous-Clustering}B the number of clusters increases until a state of random spiking is reached where each spike train forms its own cluster. As exemplified by four different time instants in each figure, this varying clustering structure is correctly reflected in the pairwise distance matrices of both the regular and the real-time SPIKE-distance although the latter only uses information from past spiking.

Both methods can not only distinguish different clusters instantaneously but are also sensitive to the detailed structure within the clusters. An example can be seen in Fig. \ref{fig:Fig5-Instantaneous-Clustering}A for the $4$th spike train within the second $500$ ms interval (see arrows). The two methods are quite similar in the first two columns, but they differ considerably in the third and fourth column. In the third column, while for the regular SPIKE-distance it does not matter whether the time instant is right within a spiking event or in between two events (compare against the first two columns), the real-time variant clearly separates neurons depending on whether they have already fired or not (cf. Fig. \ref{fig:Fig4-Not-Spurious}). In the fourth column, in contrast to the regular SPIKE-distance, the real-time SPIKE-distance is not yet aware of the irregular cluster affiliation in the last $500$ ms interval.

The main difference between the two measures is clearly visible in the last column where the regular SPIKE-distance averages over past and future behavior and thus superimposes the checkered pattern of the third interval with the more disordered clustering of the last interval. This last interval is not yet relevant for the real-time variant, which only reflects the checkered pattern of the past interval.

Similar differences can be observed in the second and third interval of Fig. \ref{fig:Fig5-Instantaneous-Clustering}B. The matrices for the regular SPIKE-distance are quite symmetric with respect to the dissimilarities to the different clusters since the increasing distances between the preceding spikes are balanced by the decreasing distances to the following spikes. In contrast, the real-time SPIKE-distance reflects the increasing distance between the preceding spikes only. Thus, although the spike time differences between adjacent groups are similar, due to the normalization lower dissimilarities are obtained for groups of spike trains whose spikes are further in the past.

\begin{figure}
    \includegraphics[width=85mm]{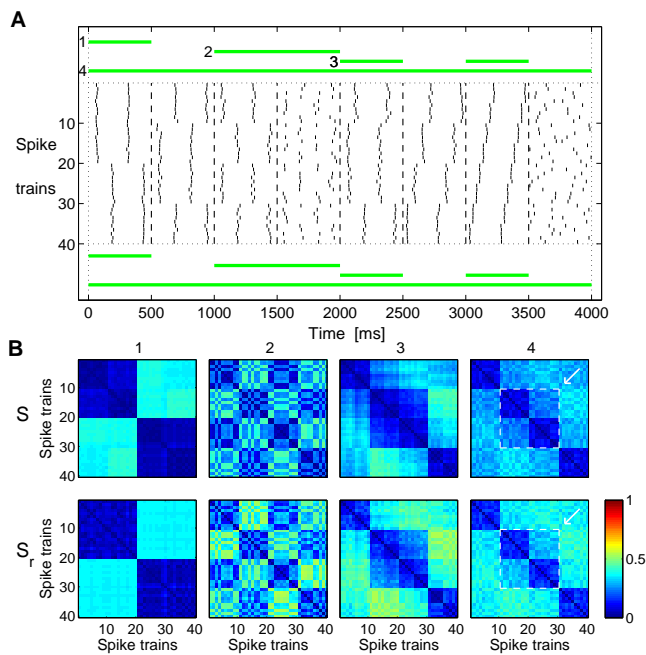}
    \caption{\abb\label{fig:Fig6-Clustering-Selective-Averaging} Selective temporal averaging.   A. The spike trains are concatenated from the two examples in Fig. \ref{fig:Fig5-Instantaneous-Clustering}.   B. Each column (from left to right) depicts the selective temporal average over the intervals marked by the horizontal lines in A (from top to bottom). Arrows indicate features described in the text.}
\end{figure}

So far we have shown clustering matrices obtained at certain time instants. Since such a matrix exists for each and every time instant $t$, it is possible to selectively average over certain time intervals. These intervals do not have to be continuous, selective averaging over separated intervals is possible as well. Four examples are shown in Fig. \ref{fig:Fig6-Clustering-Selective-Averaging}, an average over one individual $500$ ms interval, averages over two consecutive as well as two separated $500$ ms intervals, and finally the average over the whole time interval. In each case a linear superposition of the individual matrices can be observed. For the whole interval (fourth column) this means that the four groups of $10$ spike trains which frequently belonged to the same cluster can still be identified (see arrow). Due to the fact that there were two $500$ ms intervals (the second and the fifth) where the second and the third spike train group formed one big cluster, these two groups are more closely related than the other groups, which is correctly reflected by the lower values of dissimilarity.

Another option is triggered temporal averaging. Here the matrices are averaged over certain trigger time instants only. The idea is to check whether this triggered temporal average is significantly different from the global average since this would indicate that something peculiar is happening at these trigger instants. These trigger times can either be obtained from internal conditions (such as the spike times of a certain spike train) or from external influences (such as the occurrence of certain features in a stimulus).

\begin{figure}
    \includegraphics[width=85mm]{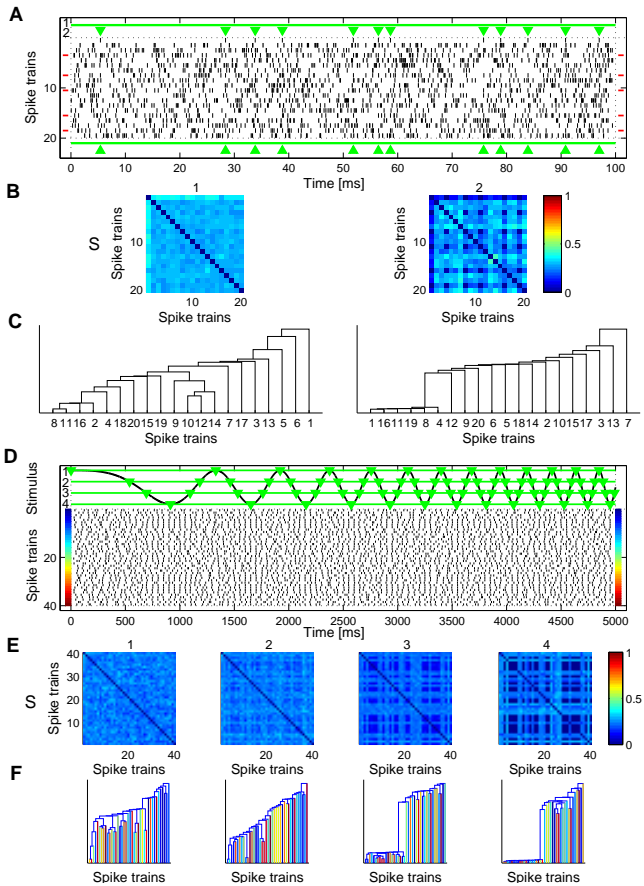}
    \caption{\abb\label{fig:Fig7-Triggered-Averaging} Triggered temporal averaging.   A-C. Internal triggering.   A. Poisson spike trains with superimposed firing patterns: Five spike trains ($\# 4, 8, 11, 16, 19$, marked by horizontal red lines) follow the spiking of the first spike train (with small amounts of jitter). Below the horizontal line denoting the overall average, triangles mark the spike times of the first spike train which are used as trigger instants.   B. Dissimilarity matrices (for the regular SPIKE-distance only): Selective averaging over the whole interval (left) and temporal averaging over the instantaneous matrices at all trigger instants (right).   C. Hierarchical cluster trees (dendrograms) obtained from the dissimilarity matrices in B.   D-F. External triggering.     D. Artificially generated spike trains. While one half of the neurons are noisy, the synchrony of the other half is modulated via a chirp-like external stimulus (shown on top). These non-periodically varying levels of synchrony can be traced by triggering on time instants with common stimulus amplitude (marked by horizontal green lines and green triangles).     E. Dissimilarity matrices. Each column (from left to right) depicts the externally triggered temporal average for decreasing amplitudes of the chirp function (from top to bottom).     F. Corresponding dendrograms. As the emerging spike train cluster shows, lower stimulus amplitudes leads to increased levels of synchrony.}
\end{figure}

An example of internal triggering can be seen in Fig. \ref{fig:Fig7-Triggered-Averaging}A-C. In this artificially generated setup there are $20$ simultaneously recorded neurons and almost all of them fire independently from each other following a Poisson statistic (Fig. \ref{fig:Fig7-Triggered-Averaging}A). The exception is the first neuron which fires at a lower rate and is assumed to have a strong excitatory synaptic coupling to five of the other neurons (\# 4, 8, 11, 16, 19). Correspondingly, the spike trains of these neurons were modified such that they contained slightly lagged and jittered copies of the spikes of the first spike train in addition to spikes generated independently. This represents a situation in which each spike in the first spike train causes (triggers) a spike in these spike trains but there are also other spikes (which might have been caused by different neurons that were not recorded).

The task is to identify these five neurons. This is very difficult via visual inspection, and also the overall temporal average is unable to do so (Fig. \ref{fig:Fig7-Triggered-Averaging}B, left column). However, these neurons can be identified by averaging over the the pairwise instantaneous values obtained for the spike times of the first spike train (the internal trigger instants) only. The resulting dissimilarity matrix shown in the right column of Fig. 7B includes an irregular grid of very small distance values.

Another representation of dissimilarity matrices are hierarchical cluster trees (dendrograms, Fig. \ref{fig:Fig7-Triggered-Averaging}C). They are constructed as follows: First, the closest pair of spike trains is identified and thereby linked by a $\sqcap$-shaped line, where the height of the connection measures the mutual distance. These two time series are merged into a single element, and the next closest pair of elements is then identified and connected. The procedure is repeated iteratively until a single cluster remains. The implementation of this method requires introducing the distance between a pair of clusters. In the single linkage algorithm used here, this distance is defined as the minimum over all the distances between pairs of spike trains in the two clusters. In the example of Fig. \ref{fig:Fig7-Triggered-Averaging}, the dendrogram obtained from the average dissimilarity matrix (left) does not fall into separate clusters, whereas in the dendrogram of the triggered average (right) the five modified spike trains form one distinct cluster with the first spike train and can thus easily be identified.

In contrast to internal triggering, externally triggered averaging allows certain (external) stimulus features to be related with spike train synchrony and might, thus, be a promising tool for the investigation of neuronal coding. While the example for internal triggering assumes a simultaneously recorded population of neurons, in the setup of the external triggering example (Fig. \ref{fig:Fig7-Triggered-Averaging}D-F) just one neuron is recorded for repeated stimulation with the same stimulus. This stimulus is a chirp-function, representative of a non-periodic time-varying stimulus. It is assumed that the neuron is sensitive to negative amplitudes and accordingly it exhibits higher (lower) reliability for local minima (maxima) of the chirp function (Fig. \ref{fig:Fig7-Triggered-Averaging}D). However, in order to better illustrate the gradual increase in clustering half the trials were left unaffected from the stimulus.

As the amplitude of the chirp function varies so does the spike train synchrony of half of the trials. Due to the non-periodicity of the stimulus this is quite difficult to detect. Detection is facilitated by externally triggered averaging where the triggering is performed on common stimulus features (in this example the amplitude of the chirp function). As can be seen in the dissimilarity matrices (Fig. \ref{fig:Fig7-Triggered-Averaging}E) and even better in the dendrogram (Fig. \ref{fig:Fig7-Triggered-Averaging}F), a decrease of this stimulus amplitude leads to the emergence of a spike train cluster consisting of the modulated spike trains which indicates their increase in reliability.

\begin{figure}
    \includegraphics[width=85mm]{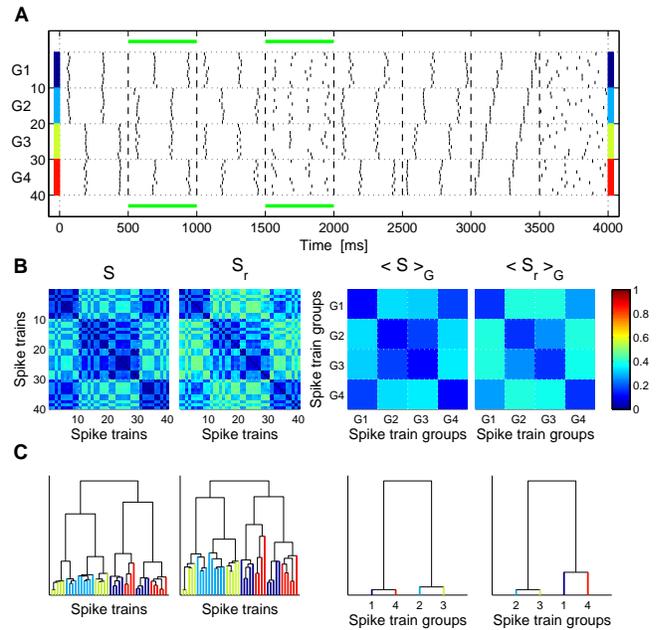}
    \caption{\abb\label{fig:Fig8-Movie-Screenshot} Screenshot from a movie (see Supplementary Material)   A. Artificially generated spike trains.   B. Dissimilarity matrices obtained by averaging over two separate time intervals for both the regular and the real-time SPIKE-distance as well as their averages over subgroups of spike trains (denoted by $<•>_G$).   C. Corresponding dendrograms.}
\end{figure}

In the supplementary material we present a movie which uses the artificially generated spike trains from Figs. \ref{fig:Fig5-Instantaneous-Clustering} - \ref{fig:Fig6-Clustering-Selective-Averaging} and includes instantaneous clustering, selective temporal averaging of individual or combined intervals, several examples of triggered averaging as well as the corresponding dendrograms. As can be seen in the screenshot of the movie shown in Fig. \ref{fig:Fig8-Movie-Screenshot}, we added one more feature, the comparison of spike train groups, where the spike trains are manually assigned to subgroups, and a block matrix (and the corresponding dendrogram) is obtained by averaging over the respective submatrices of the original dissimilarity matrix. For both distances these spatial averages over groups of spike train pairs are denoted by $<•>_G$.

\subsection{\label{ss:Application-Epilepsy} Application to single-unit recordings from epilepsy patients}

In all previous examples we have used artificially generated spike trains for which the relative levels of spike train synchrony were known and could serve as validation for the spike train distances. Here we present an exemplary application of both the SPIKE-distance and its real-time variant to field data for which no a-priori knowledge is available. As field data we chose recordings of neuronal spiking from the human medial temporal lobe. These recordings were performed at the University of Bonn in epilepsy patients undergoing seizure monitoring prior to epilepsy surgery. For a description of the data refer to Appendix \ref{App-s:Data-Single-unit-recordings-from-epilepsy-patients}.

\begin{figure}
    \includegraphics[width=85mm]{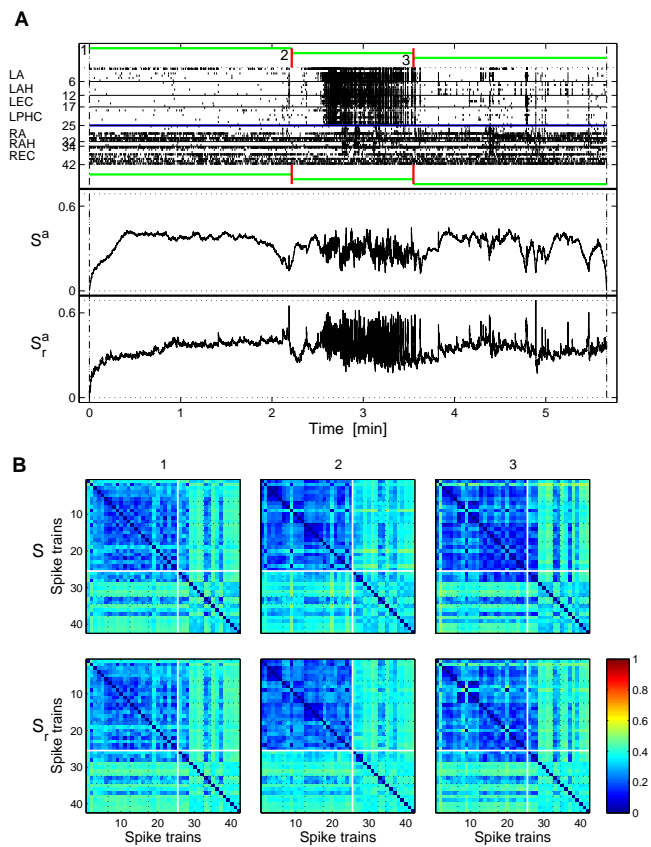}
    \caption{\abb\label{fig:Fig9-Spike-Train-Seizure} Exemplary application of the SPIKE-distance and its real-time variant to single- and multi-unit recordings from an epilepsy patient (whose epileptic focus was in the left hippocampal formation) before, during, and after an epileptic seizure.   A. Recorded spike trains and the two dissimilarity profiles (which start at zero due to the auxiliary spikes at time $t = 0$) as well as their average values for the periods before, during, and after the epileptic seizure. Seizure onset and offset are marked by vertical lines. Different brain regions (hemispheres) are separated by (thick) horizontal lines. LA=left amygdala; LAH=left anterior hippocampus; LEC=left entorhinal cortex; LPHC=left parahippocampal cortex; and accordingly for the right (R) hemisphere. No units were recorded from right parahippocampal cortex.   B. Selective temporal averaging. Each column (from left to right) depicts the selective average over the time interval marked by a horizontal line (from top to bottom) in subplot A (before, during, and after the epileptic seizure). The dashed (solid) lines separate values obtained for different brain regions (hemispheres).}
\end{figure}

In Fig. \ref{fig:Fig9-Spike-Train-Seizure}A we show the spike trains recorded from $42$ single and multi-units of an epilepsy patient during an epoch which contained an epileptic seizure. For this particular patient the epileptic focus was later confirmed to lie in the hippocampal formation of the left brain hemisphere. In this example the SPIKE-distance and its real-time variant exhibit a rather limited amount of variability before and after the epileptic seizure, while during the seizure both distances exhibit strong fluctuations, particularly during the second part of the seizure when the local field potentials exhibit high-amplitude rhythmic activity (Supplementary Fig. 1). Remarkably, both distances show a pronounced drop at the beginning of the seizure, at a time when only subtle changes are discernible in the continuous local field potential. This could be an indication that an increased level of synchrony among a population of neurons plays an important role in the generation of seizure activity.

Fig. \ref{fig:Fig9-Spike-Train-Seizure}B shows the dissimilarity matrices for both spike train distances averaged over the periods before, during, and after the seizure. These matrices reflect the relationships between all the neurons recorded from different regions of both temporal lobes. In the non-focal hemisphere, i.e. the side of the brain that does not contain the epileptic focus, neurons 26 to 42 exhibit a constant pattern of synchrony before, during, and after the seizure. In the focal hemisphere the general level of spike distance is lower during the seizure than before or afterwards, indicating an elevated level of neuronal synchrony during the seizure. The spike train distances between the two brain hemispheres (upper right and lower left quadrants) show a relatively high level of synchrony for the sparsely firing neurons ($\#26,27,28,33,34,37$) that is substantially diminished during the seizure.

These findings are in line with standard theories on seizure dynamics \citep{Engel08}. The spike train distances thus appear to be suitable measures to describe the mechanisms of seizure generation, propagation, and termination at the neuronal level. Specifically they provide an opportunity to extend our mechanistic understanding of spatio-temporal seizure dynamics by elucidating the functional role of synchronization and de-synchronization processes. A separate analysis of spike train synchrony for different regions of the focal and nonfocal medial temporal lobe may provide additional insights about the evolution of a seizure (cf. \citet{Schevon12}). The real-time variant could furthermore be integrated into prospective algorithms aimed, e.g., at an early online detection of seizure occurrence \citep{Jouny11}.

\subsection{\label{ss:Application-EEG} Application to continuous data}

To our knowledge, there are no time series analyses methods that allow to reliably track instantaneous synchrony in continuous data, i.e., to map their local dissimilarity to one value for each time instant, either once the complete segment is available for analysis or online in real-time. Now that we have provided such a method for discrete data, the question arises whether it is possible to extend their applicability to continuous data. Thus, in the third application we use the SPIKE-distance and its real-time variant to measure time-resolved dissimilarity in continuous signals. For this purpose, a high temporal resolution is once again beneficial since changes in synchronization can occur on very small time scales. Examples include the transition to seizure observable in the EEG of epilepsy patients \citep{LopesDaSilva03}.

The principal idea is to first transform the continuous time series into one or more sequences of discrete events which are chosen to capture the most relevant characteristics of the data. In the case of neuronal membrane potentials, these are the spikes. Under the assumption that both the shape of the spike and the background activity carry minimal information, neuronal responses are reduced to a spike train in which the only information maintained is the timing of the individual spikes. For the rather oscillatory EEG data that we analyze here, each continuous time series is transformed into two separate sequences of local maxima and local minima; similar choices were made e.g. in \citet{QuianQuiroga02b} and \citet{Kugiumtzis04}. The SPIKE-distance is applied to both kinds of sequences, and the two resulting dissimilarity profiles are averaged. As before, the temporal average of this combined profile yields the SPIKE-distance. For this type of data a causal calculation is also possible, the procedure is the same as used above for the regular SPIKE-distance.

\begin{figure}
    \includegraphics[width=85mm]{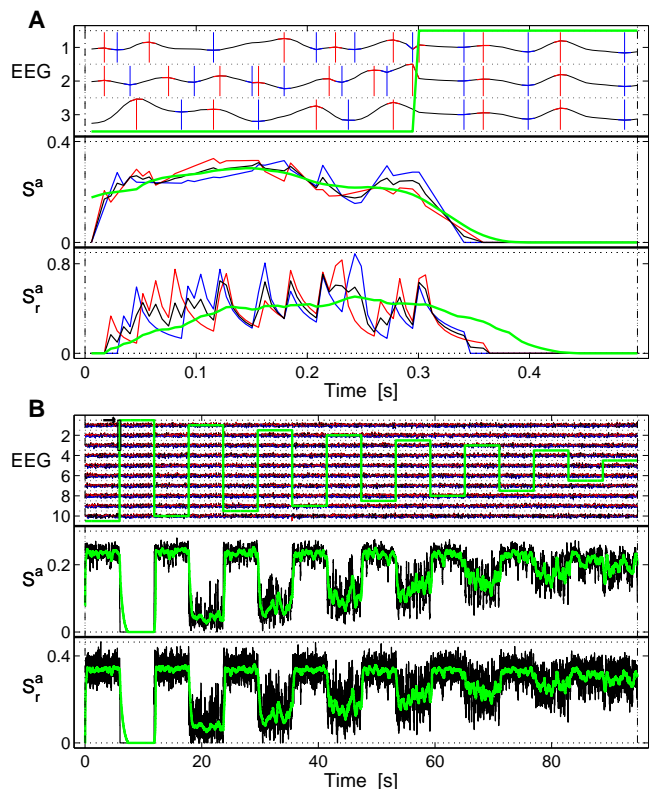}
    \caption{\abb\label{fig:Fig10-EEG-Example} Exemplary application of the averaged bivariate SPIKE-distance and its real-time variant to continuous data.   A. On top we depict three continuous EEG signals of $0.5$ s duration which exhibit a transition from being independent to being perfectly synchronous after about $0.5$ s (the mixing parameter is superimposed in green). The maxima and minima of the EEG signals are marked in red and blue, respectively, and the same colors are used below for the dissimilarity profiles obtained from applying the two SPIKE-distances to the respective event time series only. For each SPIKE-distance the black trace depicts the average over these two profiles and the green trace its moving averages.   B. On top we show ten mixed signals generated from $10$ different EEG-channels by means of an oscillating mixing parameter. For maximum mixing we obtain identical signals, while without mixing the original EEG channels are preserved. Below we depict the dissimilarity profiles of the averaged bivariate SPIKE-distance and its real-time variant, respectively, (black) as well as their moving averages (green). The excerpt used in subplot A is marked by a small black box (see arrow).}
\end{figure}

To validate the instantaneous measures of synchronization on controlled continuous data, we used data freely available on the internet via the EEG time series download page (www.meb.uni-bonn.de/epileptologie/science/physik/ eegdata.html) of the Department of Epileptology at the University of Bonn, Germany \citep{Andrzejak01b}. We randomly selected $N = 10$ independent channels (sampling rate $173.16$ Hz) and then introduced a time-dependent mixing parameter $m$. For $m = 0$ the original independent channels are maintained whereas for $m = 1/N$ all channels are exactly identical. Intermediate values of $m$ interpolate linearly between these two extremes. Although not relevant for the event detection, the variance, which due to the averaging decreases for small $m$, is adjusted to maintain the appearance of a regular EEG signal.

In Fig. \ref{fig:Fig10-EEG-Example}A we show an illustration of the application of the averaged bivariate SPIKE-distance and its real-time variant to continuous EEG data using just $3$ channels over a short interval of time which includes a transition from independent channels to perfectly synchronous channels. This transition is traced by both measures. Fig. \ref{fig:Fig10-EEG-Example}B depicts the dissimilarity profiles of the two SPIKE-distances for alternating piecewise constant variations of the mixing parameter which converges stepwise towards an intermediate level. Both measures are capable of monitoring these transitions in the level of synchronization. Starting from zero values for identical synchronization and high values for independent channels, two gradual transitions can be observed. However, as the different gradients show, it is easier to trace deviations from identical signals than deviations from independent signals.

%
\section{\label{s:Discussion} Discussion}

There are three different contributions in this study. We have improved the dissimilarity profile of the SPIKE-distance \citep{Kreuz11} by eliminating the spurious high values that were previously obtained for event-like firing patterns, we have added a variant of the SPIKE-distance which is causal and allows real-time calculation, and, finally, we have extended the applicability of both of these variants to continuous data (such as EEG).

The instantaneous reliability obtained by the elimination of the spurious high values has very important consequences. In addition to the improved capability to track the overall level of synchronization within a group of two or more spike trains (see Figs. \ref{fig:Fig2-Bi-Multi-Comparison} and \ref{fig:Fig3-Bi-Multi-Comparison-Realtime-Comp}), it is now possible to visualize instantaneous clustering (see Fig. \ref{fig:Fig5-Instantaneous-Clustering} and supplementary movie), i.e., to represent evolving patterns of (dis)similarity in multiple spike trains either as instantaneous matrices of pairwise spike train dissimilarities or as hierarchical cluster trees (dendrograms). Since such matrices and dendrograms exist for each and every time instant t, it is possible to selectively average over certain (continuous or non-continuous) time intervals (Fig. \ref{fig:Fig6-Clustering-Selective-Averaging}) which do not have to have the same length (see also Fig. \ref{fig:Fig9-Spike-Train-Seizure}). In real data these intervals could be chosen to correspond to different external conditions such as normal vs. pathological, asleep vs. awake, target vs. non-target stimulus, or presence/absence of a certain channel blocker. The fact that there are no limits to the temporal resolution allows further analyses such as internally or externally triggered temporal averaging (see Fig. \ref{fig:Fig7-Triggered-Averaging}). In real multi-neuron data, internal triggering might help to uncover certain structures and patterns in neural networks or to detect converging or diverging patterns of firing propagation. External triggering might be used to address questions of neuronal coding, e.g., it could be used to evaluate the influence of localized stimulus features on the reliability of real neurons under repeated stimulation. Finally, another possibility is spatial averaging over spike train groups (see Fig. \ref{fig:Fig8-Movie-Screenshot}). In applications to real data, these groups could be different neuronal populations or responses to different stimuli, depending on whether the spike trains were recorded simultaneously or successively.

The SPIKE-distance appears to provide a reliable time-resolved measure of spike train similarity without the need to optimize a time scale parameter. This extends the range of possible neurophysiological applications of spike train similarity measures. In the past, methods aimed at measuring spike train similarity have been applied in very controlled situations, typically protocols where the animal is anaesthetized and the same stimulus is repeated over multiple trials. With the SPIKE-distance, it is envisaged that similarity based approaches could be extended to experiments with awake behaving animals: the identical trials typically required to select an appropriate time scale parameter are not required for the SPIKE-distance and the time resolution allows the similarity to be monitored during behavior or in response to complex evolving stimuli; for example, the extent to which neurons in a population synchronize in response to stimuli could be analysed.

In cases where the complete spike trains are known, the regular SPIKE-distance compares favorably against the real-time SPIKE-distance. It is locally more reliable because it has information about both the past and the future at its disposal. For the real-time SPIKE-distance the lack of information about spikes that occur in the future leads to high values during reliable spiking events. While these values are not spurious, they are not really informative since they reflect local uncertainty which can easily be resolved by incorporating all the information available. However, in situations which demand an online monitoring the real-time SPIKE-distance can be relied upon as we could show both for simulated and field data. A good example for the latter is Fig. \ref{fig:Fig9-Spike-Train-Seizure}B in which even with only half the information available the results achieved were very similar to the ones for the regular SPIKE-distance (compare the upper and lower row).

The capability to track the synchrony between two or more spike trains (or continuous recordings) opens up new possibilities for several important applications. Synchronization has been hypothesized to play a pivotal role in neuronal coding \citep{Miller08}, and thus a real-time tracker of spike train synchrony could be an essential tool for a rapid online decoding with brain-machine interfaces in order to control prosthetic devices \citep{Hochberg06, Mulliken08, Sanchez08}. In epilepsy patients, monitoring the spiking activity of large ensembles of single neurons could lead to a better understanding of the mechanisms of seizure generation. Furthermore, if neuronal spiking patterns turn out to be specific predictors of seizure occurrence as reported by a recent study \citep{Truccolo11}, the real-time SPIKE-distance could be a viable tool for the implementation of a prospective seizure prediction algorithm \citep{Mormann07}.

In a similar manner the SPIKE-distance and its real-time variant can be applied to monitor synchrony in continuous data (which are first transformed into discrete data, see Fig. \ref{fig:Fig10-EEG-Example}). Even the analysis of mixed continuous-discrete signals is possible (see also \citeauthor{Andrzejak11}, \citeyear{Andrzejak11}). A potential application could be the combined analysis of discrete spike trains and continuous neuronal oscillations. In particular, it would be interesting to investigate neuronal synchronization patterns in dependence of the phase of the local field potential, a scenario reminiscent of the 'neuronal communication through neuronal coherence' scenario \citep{Fries05}.

Note that there is a conceptual analogy between the improved definition of the SPIKE-distance and the similarity measure proposed by \citeauthor{Hunter03} (\citeyear{Hunter03}) which basically sums the exponentially weighted distances from each spike to its nearest spike in the other spike train. However, the main difference, apart from the additional exponential weighting and the parameter (the decay constant of the exponential), is that the calculation is not done in a time-resolved manner by means of a local weighting of the terms of the four corner spikes. Instead, the term for each spike is considered just once. The SPIKE-distance is a rather elementary measure that can be regarded as complementary to cross correlation. Both measures rely on differences in spike timing. However, while the latter estimates spike synchrony in dependence on a time lag but is not sensitive to instantaneous synchrony, the SPIKE-distance estimates instantaneous synchrony but is not sensitive to time lags (these should be eliminated beforehand).

Here we provide an overview of the main properties of both the SPIKE-distance and its real-time variant:

\vspace{3 mm}

\begin{itemize}\itemsep4pt

  \item[$\bullet$] Straightforward extension to multiple spike trains and normalization

        The SPIKE-distance directly addresses the lack of approaches to analyze multiple spike train data \citep{Brown04}. It can readily be applied to very large datasets (some of the datasets analyzed contained over $100$ spike trains with a total of almost $250000$ spikes). Both the bivariate and averaged bivariate distances are normalized between zero and one where zero is obtained for identical spike trains only. The same limits hold for the underlying dissimilarity profiles. However, there is a difference regarding the range of values obtained. For the piecewise constant ISI-distance the dissimilarity profile and the distance itself cover the same range. In particular, the distance value can come arbitrarily close to the maximum value of one. For the piecewise linear SPIKE-distance this is not the case. Its value is in most cases much smaller than the maximum instantaneous value and typically below $0.5$. This is in line with the observation that for Poisson spike trains of equal rate the expectation values for the ISI- and the SPIKE-distance are $0.5$ and $0.295$, respectively.

  \item[$\bullet$] Different levels of information reduction

        For the SPIKE-distance there are three different levels of information reduction. The starting point is the most detailed representation in which one instantaneous value is obtained for each pair of spike trains. The resulting matrix of size 'number of sampled time instants' $\times$ 'squared number of spike trains' (i.e. $\# (t_n) N^2$) can be appreciated best in a movie; an example can be found in the supplementary material. For the first step of information reduction there are two possible averages: the average over spike train pairs and the temporal average. These commute, either can be performed first. By averaging instantaneously over all pairs of spike trains, a dissimilarity profile for the whole population (e.g. Fig. \ref{fig:Fig2-Bi-Multi-Comparison}) is obtained whereas the temporal averaging leads to a bivariate distance matrix (e.g. Fig. \ref{fig:Fig5-Instantaneous-Clustering}). Finally, in both cases the second average leads to one distance value which describes the overall level of synchrony for a group of spike trains over a given time interval. Note that here we restricted the analysis to average values, but also the application of higher order statistics (such as the variance) is conceivable.

        Depending on the application in mind, the appropriate representation can be chosen. Mapping the similarity of a whole population onto one single value might allow for an easier comparison but one might lose too much information since high and low spike time differences at different time instants or for different pairs of spike trains might cancel each other out, leading to a value that could also be obtained for a constant intermediate level of similarity. This is one example where a higher-order statistics such as the variance could be useful.

  \item[$\bullet$] High temporal resolution: Reliance on local information, then global averaging

        The SPIKE-distance relies on instantaneous values which only take into account local information from preceding and - with the exception of the real-time SPIKE-distance - following spikes. Temporal averaging (Eq. \ref{eq:Temporal-Average}) then yields a more global picture by means of a single-value distance. The fact that the averaging window can be chosen arbitrarily allows for easy comparisons between the levels of spike train synchrony in different time intervals without the need of recalculations. The high temporal resolution which renders a sliding-window analysis obsolete compares them favorably to other widespread measures of spike train variability such as the Fano factor.

  \item[$\bullet$] Dependence on spike train of origin

        The SPIKE-distance takes into account the spike train of origin for each individual spike and neither of them is invariant to shuffling spikes among the spike trains. This is in contrast to measures that act on the pooled spike train, such as all measures based on the Peri-Stimulus Time Histogram (PSTH), which yield the same value regardless of how spikes are distributed among the different spike trains and could possibly fail to distinguish between qualitatively different behaviors such as high reliability spiking and chain-fire bursting (cf. \citeauthor{Kreuz11}, \citeyear{Kreuz11}).

  \item[$\bullet$] Computational efficiency

        The SPIKE-distance is based on simple differences and ratios which furthermore can easily be parallelized. For a large set of very long spike trains for which computer memory might be a concern, the dissimilarity profiles for successive intervals can be calculated sequentially.

  \item[$\bullet$] Time-scale independence and absence of parameters

        The SPIKE-distance is invariant to stretching and compressing of the spike trains. It is also time-scale adaptive since the information used at each time instant is not contained within a window of fixed size but within a time interval whose size depends on the local rates of the spike trains. In contrast to time-scale dependent measures such as the Victor-Purpura \citep{Victor96} or the van Rossum distance \citep{VanRossum01}, parameter-free single-valued methods give a more objective and comparable estimate of neuronal variability. Other drawbacks of time-scale dependent measures are the computational cost and the time and effort that is needed to find the right parameter. Moreover, it is not at all guaranteed that there exists a right parameter.

        One of the main arguments for the use of time-scale-dependent measures of spike train (dis)similarity is their potential insight into the precision of the neuronal code \citep{Victor96}. This argument has recently been reevaluated in \citet{Chicharro11} for transient constant and time-varying stimuli, respectively. According to this study the optimal time-scale obtained from a spike train discrimination analysis is far from being conclusive. Rather it results in a non-trivial way from the interplay of many different factors such as the distribution of the information contained in different parts of the response and the degree of redundancy between them.

  \item[$\bullet$] SPIKE-distance is complementary to the ISI-Distance

        The SPIKE-distance and its real-time variant share many properties with the ISI-distance (see Appendix A) but there are also a few essential differences. The ISI-distance (see Appendix \ref{App-s:Bivariate-ISI-Distance}) is based on interspike intervals and quantifies covariations in the local firing rate, while the SPIKE-distance tracks synchrony mediated by spike timing. This does not mean that the ISI-distance is sensitive to rate coding and the SPIKE-distance sensitive to temporal coding. Rather, it is the relative timing of interspike intervals and spikes, respectively, that matters.

        Another difference is the range of values obtained. For the ISI-distance the piecewise constant dissimilarity profile and the distance itself cover the same range. In particular, the distance value can come arbitrarily close to the maximum value of one. For the SPIKE-distance this is not the case. Only the piecewise linear dissimilarity profile can cover the whole range of values whereas the value of the SPIKE-distance always seems to be below $0.5$. For Poisson spike trains of equal rate the expectation values for the ISI- and the SPIKE-distance are $0.5$ and $0.295$ (numerical results), respectively.

\end{itemize}

\vspace{3 mm}

The Matlab source codes for the ISI- and the SPIKE-distance (including its real-time variant) ) as well as additional material (including the supplementary figure and the supplementary movie) can be found at www.fi.isc.cnr.it/users/thomas.kreuz/sourcecode .html.

%
\begin{appendix} \label{s:Appendix}

\section{\label{App-s:Bivariate-ISI-Distance} The ISI-distance}

While the dissimilarity profile of the SPIKE-distance is extracted from differences between the spike times of the two spike trains, the dissimilarity profile of the ISI-distance \citep{Kreuz07c, Kreuz09} is calculated as the instantaneous ratio between the interspike intervals $x_{\mathrm {ISI}}^{(1)}$ and $x_{\mathrm {ISI}}^{(2)}$ (Eq. \ref{eq:ISI}) according to:
\begin{equation} \label{eq:ISI-Ratio}
    I (t) = \begin{cases}
           x_{\mathrm {ISI}}^{(1)} (t) / x_{\mathrm {ISI}}^{(2)} (t) - 1 & {\rm if} ~~ x_{\mathrm {ISI}}^{(1)} (t) \leq x_{\mathrm {ISI}}^{(2)} (t) \cr
                      - (x_{\mathrm {ISI}}^{(2)} (t) / x_{\mathrm {ISI}}^{(1)} (t) -1)     & {\rm otherwise}.
                  \end{cases}
\end{equation}
This ISI-ratio equals $0$ for identical ISI in the two spike trains, and approaches $-1$ and $1$, respectively, if the first or the second spike train is much faster than the other. For the ISI-distance the temporal averaging analogous to Eq. \ref{eq:Temporal-Average} is performed on the absolute value of the ISI-ratio, thus both kinds of deviations are treated equally. Since the ISI-distance relies on the instantaneous ISI-values and thus requires knowledge about the following spikes, no causal real-time extension is possible. The dissimilarity profile is condensed into a distance value by means of temporal averaging analogous to Eq. \ref{eq:Temporal-Average}. In \citet{Andrzejak11} the ISI-distance has been integrated in a measure which can detect unidirectional coupling not only between spike trains but also between spike trains and time-continuous flows.


\section{\label{App-s:Data-Single-unit-recordings-from-epilepsy-patients} Field data: Single-unit recordings from epilepsy patients}

All experimental recordings were performed prior to and independently from the design of this study and conformed to the guidelines of the Medical Institutional Review Board at the University of Bonn. Electrode locations were based exclusively on clinical criteria and were verified by MRI and by computer tomography co-registered to preoperative MRI. Each electrode probe had eight high-impedance (typically $800$ to $1000$ k$\Omega$) micro-wires (Platinum/Iridium, $40$ $\mu$m diameter) protruding from its tip (Adtech Inc, Racine, MN). The differential signal from bipolar montages of the micro-wires (one wire was used as local ground) was amplified using a $128$-channel Neuralynx system (Digital Lynx $10$S, Neuralynx, Bozeman, MT), filtered between $0.1$ and $9000$ Hz, and sampled continuously at $32$ kHz for later processing and analysis. The Neuralynx headstages used had unity gain, very high input impedances (ca. $1$ T$\Omega$) and no phase shift. Spike detection and sorting was performed after band-pass filtering the signals between $300$ and $3000$ Hz \citep{QuianQuiroga04}.

\end{appendix}

\vspace{1cm}

\begin{thanks}
\section{\label{s:Acknowledgement} \textbf{Acknowledgements}}

We thank Stefano Luccioli, Antonio Politi, Alessandro Torcini, Heinz Beck, Martin Pofahl, Emily Caporello, and Timothy Q. Gentner for useful discussions. We also thank Christian E. Elger for patient access, and Volker A. Coenen for electrode implantation. TK acknowledges funding support from the European Commission through the Marie Curie Initial Training Network 'Neural Engineering Transformative Technologies (NETT)', project 289146 as well as from the Italian Ministry of Foreign Affairs regarding the activity of the Joint Italian-Israeli Laboratory on Neuroscience. CH acknowledges support from the James S McDonnell Foundation through a Scholar Award in Human Cognition. RGA acknowledges grant FIS-2010-18204 of the Spanish Ministry of Education and Science. FM acknowledges support from the Lichtenberg Program of the Volkswagen Foundation.

\end{thanks}

\bibliographystyle{elsart-harv}

\pagebreak

\vspace{1cm}


Caption Supplementary Figure: An epileptic seizure recorded with microwires located in different regions of the medial temporal lobe. Some recording channels (e.g., LA3) are missing because wires were broken over the course of the EEG-video monitoring. Red vertical lines denote seizure onset. Seizure activity remained confined to the focal (left) hemisphere. Note that the neuronal spiking activity shown in Fig. \ref{fig:Fig9-Spike-Train-Seizure} was extracted from the same signals after highpass filtering. LA=left amygdala; LAH=left anterior hippocampus; LEC=left entorhinal cortex; LPHC=left parahippocampal cortex; and accordingly for the right (R) hemisphere.

\vspace{1cm}

Caption Supplementary Movie: This movie uses the artificially generated spike trains from Figs. \ref{fig:Fig5-Instantaneous-Clustering} and \ref{fig:Fig6-Clustering-Selective-Averaging}. In the first part the instantaneous dissimilarities and the clustering dendrograms of both measures are updated as the green time bar moves from left to right. The second part comprises several examples of selective temporal averaging of individual or combined intervals as well as triggered averaging.   A. Artificially generated spike trains.   B. Dissimilarity matrices obtained by averaging over two separate time intervals for both the regular and the real-time SPIKE-distance as well as their averages over subgroups of spike trains.   C. Corresponding dendrograms.

\end{document}